\documentclass[12pt,psfig]{article}
\usepackage{amssymb}
\usepackage{amsmath,amsthm}
\usepackage{amsfonts}
\usepackage{graphicx}
\usepackage{graphics}
\numberwithin{equation}{section}

\usepackage{hyperref}

\begin{document}
\begin{center}\Large\textbf{Fractional-Wrapped 
Branes with Rotation, Linear Motion and Background Fields}
\end{center}
\vspace{0.75cm}
\begin{center}{\large Elham Maghsoodi and \large Davoud
Kamani}\end{center}
\begin{center}
\textsl{\small{Physics Department, Amirkabir University of
Technology (Tehran Polytechnic)\\
P.O.Box: 15875-4413, Tehran, Iran\\
e-mails: el.maghsoodi@aut.ac.ir , kamani@aut.ac.ir \\}}
\end{center}
\vspace{0.5cm}

\begin{abstract}

We obtain two boundary states corresponding to the
two folds of a fractional-wrapped D$p$-brane, i.e. the twisted  
version under the orbifold $\mathbb{C}^{2}/\mathbb{Z}_{2}$  
and the untwisted version. The brane has rotation and 
linear motion, in the presence of the following background fields: 
the Kalb-Ramond tensor, a $U(1)$ internal gauge potential and 
a tachyon field. The rotation and linear motion 
are inside the volume of the brane.
The brane lives in the $d$-dimensional
spacetime, with the orbifold-toroidal 
structure $T^n \times \mathbb{R}^{1, d-n-5} 
\times\mathbb{C}^{2}/\mathbb{Z}_{2}$ in the twisted sector.
Using these boundary states we calculate the 
interaction amplitude of two parallel 
fractional D$p$-branes with the foregoing setup.
Various properties of this amplitude such 
as the long-range behavior will be analyzed.

\end{abstract}

{\it PACS numbers}: 11.25.Uv; 11.25.-w

\textsl{Keywords}: 
Fractional-Wrapped D-brane;
Rotation; Linear motion; 
Background fields; 
Orbifold-Toroidal spacetime; 
Twisted sector;
Untwisted sector; 
Interaction. 
\newpage
\section{Introduction} 

By using the boundary state formalism all
properties of the D-branes can be extracted. 
In this formalism a D-brane can be completely represented in
terms of all closed string states, 
internal fields, tension and dynamical variables of the brane. Hence,   
a D-brane appears as a source for 
emitting (absorbing) all closed string states.
The D-branes interaction is obtained
by overlap of two boundary states, associated with the 
branes, through the closed string propagator.
Thus, this adequate formalism has been applied for various 
configurations of the D-branes \cite{1}-\cite{21}. 

Among the different configurations of branes 
the setups with fractional D-branes have some 
appealing behaviors \cite{17}-\cite{24}. 
The fractional branes appear 
in the various parts of string and M- theories.
For example, they are useful tools for demonstrating the 
gauge/gravity correspondence \cite{24}, and the 
dynamical fractional branes prepare an 
explicit starting point for defining 
Matrix theory \cite{25, 26}. 
On the other hand, we have the compactified D-branes which 
have a considerable application in string theory. 
Besides, there are D-branes with background 
fields which possess various interesting properties.
For example, these fields drastically control the 
interactions of the branes \cite{8}-\cite{15}, and they
influence the emitted and absorbed closed
strings by the branes. 
The fractional branes, wrapped branes and the background fields
motivated us to study a configuration of the 
dynamical fractional-wrapped branes with 
background fields. 

In this paper we use the method of boundary state to 
obtain the interaction amplitude between two parallel 
fractional-wrapped bosonic D$p$-branes with 
background fields and dynamics. 
We introduce the background field $B_{\mu\nu}$, 
internal $U(1)$ gauge potentials   
and internal open string tachyon fields in 
the worldvolumes of the branes.
In addition, the branes of our setup are dynamical,
i.e. they rotate and move within their volumes.
For the background spacetime in the twisted 
sector $\mathcal{T}$ we shall apply the following topological structure 
\begin{eqnarray}
T^n \times \mathbb{R}^{1, d-n-5} 
\times\mathbb{C}^{2}/\mathbb{Z}_{2}\;\;,\;\;
n \in \{0,1,\ldots,d-5\}.
\nonumber
\end{eqnarray}
An arbitrary torus from the set
$\{T^n|n =0,1,\ldots,d-5\}$ will be considered. 
Therefore, our configuration represents a generalized setup.
We shall demonstrate that the twisted sector  
does not contribute to the long-range force,
i.e. the interaction of the distant 
branes completely comes from the untwisted sector
$\mathcal{U}$. 

This paper is organized as follows. In Sec. 2, we compute
the boundary states corresponding to a 
rotating and moving fractional-wrapped D$p$-brane 
with background and internal fields. 
In Sec. 3.1, the interaction amplitude for two parallel 
D$p$-branes
will be acquired. In Sec. 3.2, the contribution of 
the massless states of closed string to the 
interaction amplitude will be extracted. 
Section 4 is devoted to the conclusions.

\section{The boundary states corresponding to a D$p$-brane}

We start by calculating the boundary states,
associated with a fractional-wrapped D$p$-brane.
The $d$-dimensional background spacetime 
contains a toroidal compact part, 
and for the twisted sector includes a non-compact orbifold part 
$\mathbb{C}^{2}/\mathbb{Z}_{2}$. The $\mathbb{Z}_2$ group
acts on the orbifold directions $\{x^a|a= d-4, d-3, d-2, d-1\}$.
We begin with the string action 
\begin{eqnarray}
S=&-&\frac{1}{4\pi\alpha'}\int_{\Sigma} d^2\sigma
\left(\sqrt{-g} g^{ab}G_{\mu\nu} \partial_{a}X^\mu
\partial_b X^{\nu} +\epsilon^{ab}B_{\mu\nu}\partial_a X^{\mu}
\partial_b X^{\nu}\right)
\nonumber\\
&+&\frac{1}{2\pi\alpha'}\int_{\partial\Sigma}d\sigma\left(A_{\alpha}
\partial_{\sigma}X^{\alpha}
+\omega_{\alpha\beta}J^{\alpha\beta}_{\tau}
+T^2\right( X^{\alpha}\left) \right)~,
\end{eqnarray}
where $\alpha , \beta \in \{0,1,\ldots,p\}$
represent the worldvolume directions of the brane,
the metrics of the worldsheet and spacetime are $g_{ab}$ 
and $G_{\mu\nu}$, $\Sigma$ indicates the worldsheet 
of closed string and $\partial\Sigma $ is its boundary. 
Here we take the flat spacetime with the signature  
$G_{\mu\nu}=\eta_{\mu\nu}={\rm diag}(-1,1,\ldots,1 )$ 
and a constant Kalb-Ramond field $B_{\mu\nu}$.
The profile of the tachyon field is chosen as
$T^2(X) = \frac{1}{2}U_{\alpha \beta}X^{\alpha}X^{\beta}$
with the constant symmetric matrix $U_{\alpha \beta}$
\cite{27, 28}.
For the internal gauge potential we chose the gauge $A_{\alpha}=
-\frac{1}{2}F_{\alpha \beta}X^{\beta}$ with 
the constant field strength. The tachyon field and gauge
potential belong to the spectrum of the open string theory,
thus, they accurately appeared as the boundary terms.
The antisymmetric constant angular velocity $\omega_{\alpha \beta}$
shows the rotation and linear motion of the brane, and 
$J^{\alpha \beta}_\tau = X^\alpha \partial_\tau X^\beta
-X^\beta \partial_\tau X^\alpha$ 
is the angular momentum density.
Note that the rotation and linear motion of the brane 
are inside the volume of the brane. 
In fact, presence of the various internal 
fields indicates some preferred alignments in the brane, 
and hence the Lorentz symmetry in the brane worldvolume
explicitly has been broken. 
We should say that adding a tachyonic mode  
generally breaks the conformal invariance, however the 
conformal boundary state can still be considered at the 
fixed points of the orbifold. For string actions 
with tachyon fields e.g. see Ref. \cite{23} and 
references therein, and also Refs. \cite{27, 29, 30, 31, 32},
in which some of them contain the resultant boundary states.

Setting the variation of this action to zero yields    
equation of motion of $X^\mu$ and the following equations 
for the boundary state 
\begin{eqnarray}
&~&\left(\mathcal{K}_{\alpha\beta} 
\partial_{\tau}X^{\beta}+\mathcal{F}_{\alpha\beta} 
\partial_\sigma X^{\beta}
+ B_{\alpha I}\partial_\sigma X^I
+ U_{\alpha \beta }
X^{\beta}\right)_{\tau=0}|B_x\rangle=0~,
\nonumber\\
&~&\left(X^I-y^I\right)_{\tau=0}|B_x\rangle=0~,
\end{eqnarray}
where $\mathcal{K}_{\alpha
\beta}=\eta_{\alpha\beta}+4\omega_{\alpha\beta}$,
and the total field strength is 
$\mathcal{F}_{\alpha\beta}=B_{\alpha\beta}-F_{\alpha\beta}$. 
The coordinates $\{x^I|I = p+1, \ldots,d-1\}$ 
show the directions which are perpendicular to the brane worldvolume,
and the parameters $\{y^I|I = p+1, \ldots,d-1\}$ 
represent the location of the brane. Combination of Eqs. (2.2) 
eliminates the third term of the first equation.
We observe that the background fields 
impose the mixed boundary conditions along the brane 
worldvolume.

The solution of the equation of motion  
for the non-orbifold directions has the form
\begin{eqnarray}
X^{\lambda}(\sigma,\tau)&=&x^{\lambda}+2\alpha'p^{\lambda}
\tau+2L^{\lambda}\sigma~
\nonumber\\
&+&\frac{i}{2}\sqrt{2\alpha'}\sum_{m\neq 0}\frac{1}{m}
\Big{(}\alpha_m^{\lambda}e^{-2im(\tau- \sigma)}
+\tilde{\alpha}_m^{\lambda}
e^{-2im(\tau+\sigma)}\Big{)}~,\;
\end{eqnarray}
where $\lambda \in \{\alpha , I\}$ for 
the untwisted sector and 
$\lambda \in \{\alpha , i\}$ for the twisted one.  
In the twisted sector 
the set $\{x^i|i=p+1,\ldots,d-5\}$
represents the non-orbifold perpendicular directions
to the brane worldvolume.
In the solution (2.3) for the non-compact coordinates,
like the time direction,
the quantity $L^{\lambda}$ identically vanishes, 
while for the circular directions there are 
\begin{eqnarray}
&~&L^{\lambda}=N^{\lambda}R^{\lambda}~,\;\;\;\;\;\; 
N^{\lambda}\in \mathbb{Z}~,
\nonumber\\
&~&p^{\lambda}=\dfrac{M^{\lambda}}{R^{\lambda}}~,
\;\;\;\;\;\;\;\;\; M^{\lambda}\in \mathbb{Z}~,
\end{eqnarray}
where $N^{\lambda}$ is the winding number 
and $M^{\lambda}$ is momentum number 
of a closed string state, and $R^{\lambda}$ 
specifies the radius of compactification 
for the compact direction $x^{\lambda}$.
Now look at the orbifold directions.
The orbifold $\mathbb{C}^{2}/\mathbb{Z}_{2}$
is non-compact, thus, its fixed points define a 
$(d-4)$-dimensional hyperplane 
at $x^a =0$. As the D$p$-brane has to sit 
on this hyperplane, and as the closed string is emitted
(absorbed) at the brane position,
the orbifold coordinates of the closed string 
possess the solution 
\begin{equation}
X^a(\sigma,\tau)=\frac{i}{2}\sqrt{2\alpha'}
\sum_{r\in\mathbb{Z}+\frac{1}{2}}
\frac{1}{r}\Big{(}\alpha_r^{a}
e^{-2ir(\tau- \sigma)}+\tilde{\alpha}_r^a
e^{-2ir(\tau+\sigma)}\Big{)}.
\end{equation}

In the twisted sector the solutions (2.3) 
and (2.5) decompose the second 
equation of (2.2) as in the following 
\begin{eqnarray}
&~&(X^i-y^i)_{\tau=0}|B\rangle^{\mathcal{T}}=0~,
\nonumber\\
&~&(X^a)_{\tau=0}|B\rangle^{\mathcal{T}}=0~.
\end{eqnarray}

By introducing Eqs. (2.3) and (2.5) into the boundary state equations we
acquire the following equations 
\begin{eqnarray}
&~&\bigg{[}\left(\mathcal{K}_{\alpha\beta}
- \mathcal{F}_{\alpha\beta}+\dfrac{i}{2m}
U_{\alpha\beta}\right)
\alpha_{m}^{\beta}+\left(\mathcal{K}_{\alpha\beta}
+ \mathcal{F}_{\alpha\beta}
-\dfrac{i}{2m}U_{\alpha\beta}\right)
\tilde{\alpha}_{-m}^{\beta}
\bigg{]}|B_{\rm osc}\rangle^{\mathcal{T} \setminus \mathcal{U}}=0,
\nonumber\\
&~&\left( 2\alpha' \mathcal{K}_{\alpha\beta}p^{\beta}
+ 2\mathcal{F}_{\alpha\beta}  
L^{\beta}+U_{\alpha\beta }x^{\beta}
\right) |B\rangle^{(0)\mathcal{T}\setminus\mathcal{U}}=0~,
\nonumber\\
&~& U_{\alpha\beta}L^{\beta}
|B\rangle^{(0)\mathcal{T}\setminus\mathcal{U}}=0,
\end{eqnarray}
for both twisted and untwisted sectors, and
\begin{eqnarray}
&~&(\alpha_{m}^{i}-\tilde{\alpha}_{-m}^{i})
|B_{\rm osc}\rangle^{\mathcal{T}}=0,
\nonumber\\
&~&(\alpha_{r}^{a}-\tilde{\alpha}_{-r}^{a})
|B_{\rm osc}\rangle^{\mathcal{T}}=0,
\nonumber\\
&~&(x^i-y^i)|B\rangle^{(0)\mathcal{T}}=0,
\nonumber\\
&~& L^{i}|B\rangle^{(0)\mathcal{T}}=0,
\end{eqnarray}
for the twisted sector, and
\begin{eqnarray}
&~&(\alpha_{m}^{I}-\tilde{\alpha}_{-m}^{I})
|B_{\rm osc}\rangle^{\mathcal{U}}=0,
\nonumber\\
&~&(x^I-y^I)|B\rangle^{(0)\mathcal{U}}=0,
\nonumber\\
&~& L^{I}|B\rangle^{(0)\mathcal{U}}=0,
\end{eqnarray}
for the untwisted sector, where we applied 
$|B_x\rangle=|B\rangle^{(0)} \otimes|B_{\rm osc}\rangle $.
Since the fractional brane has stuck at the fixed 
points of the orbifold the state $|B\rangle^{(0)}$
does not obtain any contribution 
from the orbifold directions.

According to the third equation of Eqs. (2.7) the tachyon field  
plays a crucial role for winding of closed strings
around the compact directions of the brane.
This equation implies that if the tachyon matrix 
is invertible we obviously receive the zero winding numbers 
$\{N^{\bar \alpha} =0|{\bar \alpha}=1,2,\ldots,p\}$,
and hence closed strings cannot wrap 
around the circular directions of the brane. If the 
tachyon matrix possesses null determinant the 
vector $\{L^{\bar \alpha}|{\bar \alpha}=1,2,\ldots,p\}$
can be nonzero, and therefore such wrapping of closed strings are
allowable. If the perpendicular direction 
$x^i$ (or $x^I$) is non-compact the last equation of Eqs. (2.8) 
(or (2.9)) becomes trivial, i.e. $L^i$ (or $L^I$) 
identically vanishes, 
and if $x^i$ (or $x^I$) is compact we observe 
that closed strings cannot wrap around it,
that is $N^i =0$ (or $N^I =0$).

The second equation of Eqs. (2.7) eventuates to the following 
valuable relation between the eigenvalues
\begin{eqnarray}
p^{ \alpha} =-\frac{1}{2\alpha'} 
\left[ \left(\mathcal{K}^{-1}U\right)^{ \alpha}_{\;\;{\beta}} 
x^{\beta}+ 2\left(\mathcal{K}^{-1} 
\mathcal{F}\right)^{\alpha}_{\;\;{\beta}} 
\ell^{\beta}\right] ,
\end{eqnarray}
where $\ell^{\beta}$ is eigenvalue of the operator $L^{\beta}$.
We observe that any closed string state 
(wrapped or unwrapped) has a spacetime momentum along the
worldvolume of the brane. This momentum includes two parts: 
continuous and discrete. The former 
is created by the tachyon while the latter 
originates from the Maxwell field and compactification. 
As we see this momentum is somewhat under the influence of the 
rotation and linear motion of the brane.
This nonzero momentum extremely is unlike the 
conventional case in which the closed strings are
radiated perpendicular to the brane worldvolume, 
for the conventional case e.g. see Refs. \cite{7, 33, 34}.
Thus, a peculiar potential, which is inspired by the background 
fields, the brane dynamics and compactification, 
acts on the center-of-mass positions of the 
emitted closed strings. 
If the brane directions are non-compact 
and or they are compact but the tachyon 
matrix is invertible Eq. (2.10) reduces to
\begin{eqnarray}
p^{\alpha} =-\frac{1}{2\alpha'} 
\left(\mathcal{K}^{-1}U\right)^{\alpha}_{\;\;{\beta}} 
x^{\beta}.
\end{eqnarray}

By the quantum mechanical technics, specially by 
using the commutation relations between $x^\alpha$ 
and $p^\beta$, and between $x'^\alpha$ and 
$L^\beta/\alpha'$, where $x'^\alpha =x^\alpha_L-x^\alpha_R$ 
and $L^\alpha =\alpha'(p^\alpha_L-p^\alpha_R )$,
the zero-mode part of the boundary state in the twisted
sector finds the form 
\begin{eqnarray}
|B\rangle^{(0)\mathcal{T}}&=&\frac{T_p}{2\sqrt{\det(U/2)}}
\int_{-\infty}^{\infty}
\exp\bigg{[}i\alpha' \sum_{\alpha \neq \beta}
\left(U^{-1}\mathcal{K}
+\mathcal{K}^T U^{-1}\right)_{\alpha\beta}\;
p^{\alpha}p^{\beta}
\nonumber\\
&+& \frac{i}{2}\alpha' \left(U^{-1}\mathcal{K}
+\mathcal{K}^T U^{-1}\right)_{\alpha \alpha}\;
(p^{\alpha})^2 
+ 2i\left(U^{-1}
\mathcal{F}\right)_{\alpha \beta}\;\ell^{\alpha}p^{\beta}
\bigg{]}
\nonumber\\
&\times &
\prod^{d-5}_{i=p+1} \left[\delta \left({x}^{i}-y^{i}\right)
|p^{i}_{L}=p^{i}_{R}=0 \rangle \right] \prod^{p}_{\alpha =0}
\left(|p^{\alpha}\rangle dp^{\alpha}\right)~
\label{zer}.
\end{eqnarray}
The disk partition function induces the normalization
factor $1/\sqrt{\det(U/2)}$, \cite{35, 36}. 
In the same sector, by 
using the coherent state method \cite{37}, we obtain the following 
boundary state for the closed string oscillators
\begin{eqnarray}
|B_{\rm osc}\rangle^{\mathcal{T}} 
&=&\prod_{n=1}^{\infty}[\det{M_{(n)}}]^{-1}
\exp\left[{-\sum_{m=1}^{\infty}
\left(\frac{1}{m}\alpha_{-m}^{\lambda}S_{(m)\lambda\lambda'}
\tilde{\alpha}_{-m}^{\lambda'}\right)}\right]
\nonumber\\
&\times&\exp\left[-\sum_{r=1/2}^{\infty}
\left(\frac{1}{r}\alpha_{-r}^{a}\tilde{\alpha}_{-r}^{a}\right)\right]
|0\rangle_\alpha|0\rangle_{\tilde{\alpha}}~,
\label{aos}
\end{eqnarray}
where $\lambda , \lambda' \in \{\alpha , i\}$, and 
the matrix $S_{(m)}$ is defined by
\begin{eqnarray}
S_{(m)\lambda\lambda'}&=&\left(Q_{(m)\alpha \beta} \equiv (M_{(m)}^{-1}
N_{(m)})_{\alpha\beta},-\delta_{ij}\right)~,
\nonumber\\
M_{(m)\alpha\beta}&=&\mathcal{K}
_{\alpha\beta}- \mathcal{F}_{\alpha\beta}
+\dfrac{i}{2m}U_{\alpha\beta}~,
\nonumber\\
N_{(m)\alpha\beta}&=&\mathcal{K}_{\alpha\beta}
+ \mathcal{F}_{\alpha\beta}-\dfrac{i}{2m}U_{\alpha\beta}~.
\end{eqnarray}
Expansion of the exponential parts of Eq. (2.13)
clarifies that the brane couples to the whole 
closed string spectrum in the twisted sector. 
The disk partition function gives the normalizing 
factor $\prod_{n=1}^{\infty}[\det{M_{(n)}}]^{-1}$
\cite{1, 16, 36}.
More precisely, the quadratic forms of the tachyon 
profile and rotating-moving term, accompanied by the 
gauge $A_\alpha = -\frac{1}{2}F_{\alpha\beta}X^\beta$, 
give a quadratic form to the boundary part of the action 
(2.1). Thus, there exists a Gaussian path integral, which 
induces the prefactors of Eqs. (2.12) and (2.13), and 
also the prefactors of the next Eqs. (2.15) and (2.16).

In a similar fashion, the untwisted sector $\mathcal{U}$ 
has the following boundary states for the zero-mode 
part and the oscillating part
\begin{eqnarray}
|B\rangle^{(0) \mathcal{U}}&=&\frac{T_p}{2\sqrt{\det(U/2)}}
\int_{-\infty}^{\infty}
\exp\bigg{[}i\alpha' \sum_{\alpha \neq \beta}
\left(U^{-1}\mathcal{K}
+\mathcal{K}^T U^{-1}\right)_{\alpha\beta}\;
p^{\alpha}p^{\beta}
\nonumber\\
&+& \frac{i}{2}\alpha' \left(U^{-1}\mathcal{K}
+\mathcal{K}^T U^{-1}\right)_{\alpha \alpha}\;
(p^{\alpha})^2 
+ 2i\left(U^{-1}
\mathcal{F}\right)_{\alpha \beta}\;\ell^{\alpha}p^{\beta}
\bigg{]}
\nonumber\\
&\times& \prod^{d-1}_{I=p+1} \left[\delta \left({x}^{I}-y^{I}\right)
|p^{I}_{L}=p^{I}_{R}=0 \rangle \right] \prod^{p}_{\alpha =0}
\left(|p^{\alpha}\rangle dp^{\alpha}\right)~
\label{zer},
\end{eqnarray}
\begin{eqnarray}
|B_{\rm osc}\rangle^{\mathcal{U}} 
&=&\prod_{n=1}^{\infty}[\det{M_{(n)}}]^{-1}
\exp\left[{-\sum_{m=1}^{\infty}
\left(\frac{1}{m}\alpha_{-m}^{\lambda}S_{(m)\lambda\lambda'}
\tilde{\alpha}_{-m}^{\lambda'}\right)}\right]
|0\rangle_\alpha|0\rangle_{\tilde{\alpha}}~,
\label{aos}
\end{eqnarray}
where $\lambda , \lambda' \in \{\alpha , I\}$, and 
$S_{(m)\lambda\lambda'}=\left(Q_{(m)\alpha \beta},
-\delta_{IJ}\right)$. 

For obtaining Eq. (2.15) we have used methods of quantum 
mechanics, specially the commutation relations between the 
position coordinates and their corresponding momenta, 
and for Eq. (2.16) we have applied the coherent 
state method. As expected, by setting all 
linear and angular velocities 
to zero the above boundary states reduce to the simple 
configurations of the D-branes, e.g. see Ref. \cite{12}. 
Besides, by decompactifying the compact directions and 
quenching the background fields and velocities we 
receive the simpler boundary states, 
e.g. see Refs. \cite{6, 24, 34, 38}.

Look at the first equation of Eqs. (2.7).
The coherent state method on the oscillators 
$\{\alpha^\beta_m \;,\;{\tilde \alpha}^\beta_{-m}
| m=1,2,3,\ldots\}$ 
introduces the matrix $Q_{(m)\alpha \beta}$ in Eqs. 
(2.13) and (2.16), while this method on the set
$\{{\tilde \alpha}^\beta_{m}\;,\;\alpha^\beta_{-m}
| m=1,2,3,\ldots\}$
recasts these boundary states with the matrix 
$\left([Q^{-1}_{(-m)}]^T\right)_{\alpha \beta}$.
Equality of these matrices leads to the following conditions  
\begin{eqnarray}
&~& \eta U -U\eta +4(\omega U+ U\omega ) =0 ,
\nonumber\\
&~& \eta \mathcal{F} - \mathcal{F} \eta 
+4(\omega \mathcal{F} +\mathcal{F}\omega)=0.
\end{eqnarray}
These equations are independent of the mode numbers.

Finally we shall use the following known boundary state, 
corresponding to the conformal ghost fields \cite{16, 34},
\begin{equation}
|B_{\rm gh}\rangle=\exp{\left[\sum_{m=1}^{\infty}(c_{-m}\tilde{b}_{-m}
-b_{-m} \tilde{c}_{-m})\right]}\frac{c_0+\tilde{c}_0}{2}
|q=1\rangle|\tilde{q}=1\rangle~.
\end{equation}
This state is 
independent of the orbifold projection, toroidal 
compactification, rotation and linear motion of the brane 
and the background fields.
The total boundary state in the bosonic 
string theory, for each sector, is given by
\begin{equation}
|B\rangle^{\mathcal{T}\setminus \mathcal{U}}_{\rm Total}
=|B_{\rm osc}\rangle^{\mathcal{T}\setminus \mathcal{U}} 
\otimes|B\rangle^{(0)\mathcal{T}\setminus \mathcal{U}}
\otimes|B_{\rm gh}\rangle~.
\end{equation}

Compare the boundary states (2.12), (2.13), 
(2.15) and (2.16) with the boundary states of
a bare brane, i.e. a stationary brane without 
any background and internal fields. 
This induces to define the following 
effective tension for the dressed brane 
\begin{equation}
{\mathcal{T}}_p =
\frac{T_p}{\sqrt{\det(U/2)}}
\bigg{|}\prod_{n=1}^{\infty}[\det{M_{(n)}}]^{-1}\bigg{|}.
\end{equation}

\section{Interaction between two D$p$-branes}

The interactions of the branes have appeared 
in many physical phenomena and in the main problems of physics.
For example, in the brane-world scenario these 
interactions have been introduced as the origin 
of the inflation \cite{35, 39}.
Beside, interaction and collision of two D-branes 
create a Big-Bang \cite{40}.
In addition, in the early universe these 
interactions have been considered 
for describing the radiation-dominated era.
Also there are D$p$-branes that overlap with our D3-brane,
hence, interact with it. Thus, these interactions induce  
the added gravity within our world \cite{41, 42}.
Furthermore, the branes interactions 
clarify some corners of the gauge/gravity correspondence \cite{24}.
Finally, the gravitational interaction between the branes
describes creation of the dark matter \cite{43}.
There are many other satisfactory applications of 
such interactions, e.g. see the Refs. \cite{36, 44, 45, 46}.

The interaction between two D-branes can be  
described by the 1-loop graph of an open string worldsheet 
\cite{47}-\cite{49}, or tree-level diagram of a closed
string worldsheet \cite{1}-\cite{21}.
In the second approach each brane couples to
all closed string states through its corresponding boundary state.
This is due to the fact that 
all properties of a D-brane are encoded into a boundary state.
Thus, in the closed string channel closed string is radiated 
from one brane, then propagates
toward the other brane, and finally is absorbed by the second brane.
Therefore, for acquiring the interaction amplitude of 
two D$p$-branes we should calculate the overlap of their 
corresponding boundary states via the closed
string propagator, i.e.,
\begin{eqnarray}
\mathcal{A}=\langle
B_1|D|B_2\rangle~,
\end{eqnarray}
where the total boundary states of the branes 
should be used. ``$D$'' is the closed string propagator, and 
is constructed from the closed string Hamiltonian. 
For the twisted sector the Hamiltonian is
\begin{eqnarray}
H^{\mathcal{T}} &=& H_{\rm ghost}+\alpha'p^{\lambda}p_{\lambda}
+ 2\left(\sum_{n=1}^{\infty}(\alpha_{-n}^{\lambda}
\alpha_{n\lambda}
+\tilde{\alpha}_{-n}^{\lambda}\tilde{\alpha}_{n\lambda})
+\sum_{r=1/2}^{\infty}
(\alpha_{-r}^{a}\alpha_{ra}
+\tilde{\alpha}_{-r}^{a}\tilde{\alpha}_{ra})\right)
\nonumber\\
&-& \frac{d-6}{6}~,\;\;\;\lambda \in \{\alpha , i\}.
\end{eqnarray}
For the untwisted sector there is
\begin{eqnarray}
H^{\mathcal{U}} = H_{\rm ghost}+\alpha'p^{\lambda}p_{\lambda}
+ 2\sum_{n=1}^{\infty}(\alpha_{-n}^{\lambda}
\alpha_{n\lambda}+\tilde{\alpha}_{-n}^{\lambda}\tilde{\alpha}_{n\lambda})
-d/6~,\;\;\;\lambda \in \{\alpha , I\}.
\end{eqnarray}
The difference between the ground state energies of 
the two sectors is a consequence of the
orbifold projection on the twisted sector. These 
ground state energies 
impose some significant effects in the branes interaction.

\subsection{Interaction amplitude: arbitrary distance of the branes}

According to the orbifold projection 
the total interaction amplitude has two parts: one part
from the untwisted sector and the other part 
from the twisted sector
\begin{eqnarray}
\mathcal{A}^{\rm Total}&=&\mathcal{A}^{\mathcal{T}}
+\mathcal{A}^{\mathcal{U}}~.
\end{eqnarray}
After a heavy calculation we receive the following 
amplitude for the twisted sector  
\begin{eqnarray}
\mathcal{A}^{\mathcal{T}}&=&\frac{T_p^2\alpha'V_{p+1}}{4(2\pi)^{d-p-5}}
\frac{\prod_{n=1}^{\infty}[\det(M^\dagger_{(n)1} 
M_{(n)2})]^{-1}}{\sqrt{\det{(U_1/2)}\det{(U_2/2)}}}
\int_{0}^{\infty}dt\bigg{[}e^{(d-8)t/6}\left(
\sqrt{\frac{\pi}{\alpha' t}}\right)^{d_{i_n}}
\nonumber\\
&\times& \exp\left( {-\frac{1}{4\alpha't}
\sum_{i_n}{\left(y_{1}^{i_n}-y_{2}^{i_n}\right)^2}}
\right)\prod_{i_{c}}\Theta_{3} 
\left(\dfrac{y_{1}^{i_{c}}-y_{2}^{i_{c}}}
{2\pi R_{i_{c}}} \bigg{|}
\dfrac{i\alpha' t}{\pi R_{i_{c}}^{2}}\right)
~\nonumber\\
&\times& [\det \mathcal{Z}(t)]^{-1/2}
\sum_{\{N^{\alpha_c}\}}\exp\left(2 W^{\dagger}
\mathcal{Z}(t)^{-1}W\right)
~\nonumber\\
&\times& \prod_{n=1}^\infty \bigg{(} 
\det[\mathbf{1}-Q^\dagger_{(n)1}
Q_{(n)2}e^{-4nt}]^{-1}~
\left(1- e^{-4nt}\right)^{p-d+7}
\left(1- e^{-2(2n-1)t}\right)^{-4}\bigg{)}\bigg{]},
\end{eqnarray}
where $V_{p+1}$ is the common worldvolume of the branes, and
\begin{eqnarray}
&~& W_{\alpha} = (U_1^{-1}\mathcal{F}_1)_{\beta_{c}
\alpha}\ell^{\beta_{c}}+(U_2^{-1}\mathcal{F}_2)
_{\beta_{c}\alpha}\ell^{\beta_{c}}~,
\nonumber\\
&~& {\mathcal{Z}(t)}_{\alpha\beta} = 
\begin{cases} 
2 t\alpha'\delta_{\alpha\beta}
+i \alpha'[(U_1^{-1}\mathcal{K}_1
+ \mathcal{K}^T_1 U_1^{-1})
-(U_2^{-1}\mathcal{K}_2+ \mathcal{K}^T_2 
U_2^{-1})]_{\alpha\beta},  
& \mbox{if }\alpha=\beta \\
2i \alpha'[(U_1^{-1}\mathcal{K}_1
+ \mathcal{K}^T_1 U_1^{-1})
-(U_2^{-1}\mathcal{K}_2+ \mathcal{K}^T_2 
U_2^{-1})]_{\alpha\beta}, & \mbox{if }\alpha\neq\beta.
\end{cases} 
\end{eqnarray}
Besides, we decomposed each set of the directions into
the compact and non-compact subsets, i.e.
\begin{eqnarray} 
\lbrace i = p+1 , \ldots, d-5
\rbrace=\lbrace i_{n}
\rbrace\cup\lbrace i_{c}\rbrace \;\;\;\;,\;\;\;\;
\lbrace \alpha = 0, \ldots, p
\rbrace=\lbrace 
\alpha_{n} \rbrace\cup\lbrace \alpha_{c} \rbrace,
\nonumber
\end{eqnarray}
where the index ``c'' (``n'') represents the word 
``compact'' (``non-compact''). 
Thus, $d_{i_n}$ is the dimension of the directions $\{x^{i_n}\}$.
The factor $\prod_{n=1}^\infty(1- e^{-4nt})^{p-d+7}$ 
originates from the oscillators of the non-orbifoldy perpendicular 
directions and the conformal ghosts, and the last factor 
of the last line is contribution of the orbifold directions.

The interaction amplitude in the untwisted sector is given by 
\begin{eqnarray}
\mathcal{A}^{\mathcal{U}}&=&\frac{T_p^2\alpha'V_{p+1}}{4(2\pi)^{d-p-1}}
\frac{\prod_{n=1}^{\infty}[\det(M^\dagger_{(n)1} 
M_{(n)2})]^{-1}}{\sqrt{\det{(U_1/2)}\det{(U_2/2)}}}
\int_{0}^{\infty}dt\bigg{[}e^{(d-2)t/6}\left(
\sqrt{\frac{\pi}{\alpha' t}}\right)^{d_{I_{n}}}
\nonumber\\
&\times& \exp\left( {-\frac{1}{4\alpha't}
\sum_{I_n}{\left(y_{1}^{I_n}-y_{2}^{I_n}\right)^2}}
\right)\prod_{I_{c}}\Theta_{3} 
\left(\dfrac{y_{1}^{I_{c}}-y_{2}^{I_{c}}}
{2\pi R_{I_{c}}} \bigg{|}
\dfrac{i\alpha' t}{\pi R_{I_{c}}^{2}}\right)
~\nonumber\\ 
&\times& [\det \mathcal{Z}(t)]^{-1/2}
\sum_{\{N^{\alpha_c}\}}\exp\left(2 W^{\dagger}
\mathcal{Z}(t)^{-1}W\right)
~\nonumber\\
&\times& \prod_{n=1}^\infty \bigg{(} 
\det[\mathbf{1}-Q^\dagger_{(n)1}
Q_{(n)2}e^{-4nt}]^{-1}~
\left(1- e^{-4nt}\right)^{p-d+3}\bigg{)}\bigg{]},
\end{eqnarray}
where $\lbrace I = p+1 , \ldots, d-1
\rbrace=\lbrace I_{n}
\rbrace\cup\lbrace I_{c}\rbrace $, and 
$d_{I_n}= {\rm dim}\;\{x^{I_n}\}$.
The factor $\prod_{n=1}^\infty(1- e^{-4nt})^{p-d+3}$ 
originates from the oscillators of the perpendicular 
directions and the conformal ghosts.

For computing the amplitudes (3.5) and (3.7)
we receive the factor 
$\prod_{\alpha =0}^p \langle p_1^\alpha | p_2^\alpha 
\rangle$. This implies that a nonzero interaction 
requires the equation  
\begin{eqnarray}
p_1^\alpha - p_2^\alpha =0, \;\;\;\alpha=0,1,\ldots,p.
\nonumber
\end{eqnarray}
According to Eq. (2.10) this equation eventuates to the following 
conditions 
\begin{eqnarray}
&~& \det \left(\mathcal{K}_1^{-1}U_1 - \mathcal{K}_2^{-1}U_2
\right)=0,
\nonumber\\
&~& \det \left(\mathcal{K}_1^{-1}\mathcal{F}_1 - 
\mathcal{K}_2^{-1}\mathcal{F}_2 \right)=0.
\end{eqnarray}
The conditions (2.17) and (3.8)
reduce $n+(p+1)(3p+2)/2$ parameters of the theory to
$n-2+p(p+1)/2$, where ``$n$'' is the dimension of the 
asymmetric torus $T^n$.

The second lines of the amplitudes (3.5) and (3.7) imply that  
the interaction is exponentially
damped by the square distance of the branes. 
In the last lines of these equations  
the determinants come from the 
oscillators of the string coordinates 
$\{X^\alpha\}$. The overall factors in front of the
integrals, which include the parameters
of the system, partially specify 
the strength of the interaction.

The variety of the parameters in the setup, 
i.e., the matrix elements of: the Kalb-Ramond tensor and  
field strengths and  
tachyon matrices, the linear and angular speeds of the branes,
the dimensions of the spacetime and the branes, 
the closed string 
winding and momentum numbers, the coordinates 
of the branes location, and the radii of the 
circular directions, specifies a general interaction 
amplitude $\mathcal{A}^{\rm Total}=\mathcal{A}^{\mathcal{T}}
+\mathcal{A}^{\mathcal{U}}$. 

The effects of the toroidal compactification have been gathered 
in $i_n$, $d_{i_n}$, $I_n$, $d_{I_n}$,
the Jacobi theta function $\Theta_3$
and the worldvolume vector $W_\alpha$. Thus, 
for obtaining the interaction amplitudes 
in the non-compact spacetime it is sufficient to  
exert the following replacements: $i_n \rightarrow i$, 
$d_{i_n} \rightarrow d_i=d-p-5$, 
$\Theta_3 \rightarrow 1$
and $W \rightarrow 0$ in Eq. (3.5); and $I_n \rightarrow I$, 
$d_{I_n} \rightarrow d_I=d-p-1$, $\Theta_3 \rightarrow 1$
and $W \rightarrow 0$ in Eq. (3.7).

\subsection{Interaction amplitude: large distance of the branes}

Behavior of the total interaction 
amplitude for large distances of the branes is very important. 
This prominently defines the long-range force of the theory, 
which is determined by
\begin{eqnarray}
\mathcal{A}_{\rm long-range}^{\rm Total}
&=& \mathcal{A}_{\rm long-range}^{\mathcal{T}}
+\mathcal{A}_{\rm long-range}^{\mathcal{U}}~.
\end{eqnarray}
In fact, this picks out the contributions of the closed 
string tachyon and massless states to the interaction.
For this purpose, since the states of the graviton, 
Kalb-Ramond tensor and dilaton 
have zero winding and zero momentum numbers we shall impose
$\ell^{\beta_c}=0$ for every $\beta_c$.
Besides, in the critical string theory, i.e. for the 
dimension $d=26$, we impose the limit $t\rightarrow \infty$ 
on the oscillating parts of the amplitudes (3.5) and (3.7).
Since the nature of an emitted (absorbed)
closed string is independent of the locations of the 
interacting branes the position factors in
Eqs. (3.5) and (3.7) do not change.
In this limit the contribution of all massive states, except 
the tachyon state, vanish. 

For the twisted sector the limit is
\begin{eqnarray}
&~& {\mathop{\lim }_{t\to \infty}}
e^{3t} \prod_{n=1}^\infty \bigg{(} 
\det[\mathbf{1}-Q^\dagger_{(n)1}
Q_{(n)2}e^{-4nt}]^{-1}~
\left(1- e^{-4nt}\right)^{p-d+7}
\left(1- e^{-2(2n-1)t}\right)^{-4}\bigg{)}
\nonumber\\
&~& \longrightarrow
e^{3t}+\left[21-p +{\rm Tr}\left(Q^\dagger_{(n=1)1} 
Q_{(n=1)2}\right)\right]e^{-t}.
\end{eqnarray}
Thus, the interaction amplitude 
of the distant branes, in the twisted sector, has the following form 
\begin{eqnarray}
\mathcal{A}_{\rm long-range}^{\mathcal{T}} 
&=&\frac{T_p^2\alpha'V_{{p+1}}}{4(2\pi)^{21-p}}
\frac{\prod_{n=1}^{\infty}[\det(M^\dagger_{(n)1} 
M_{(n)2})]^{-1}}{\sqrt{\det{(U_1/2)}\det{(U_2/2)}}}
\int_{0}^{\infty}dt\bigg{\{}\Big{(}
\sqrt{\frac{\pi}{\alpha' t}}\Big{)}^{d_{i_n}}
~\nonumber\\
&\times& [\det \mathcal{Z}(t)]^{-1/2}
\exp\left( {-\frac{1}{4\alpha't}
\sum_{i}{\left(y_{1}^{i_n}-y_{2}^{i_n}\right)^2}}
\right)\prod_{i_{c}}\Theta_{3} 
\left(\dfrac{y_1^{i_{c}}-y_2^{i_{c}}}{2\pi R_{i_{c}}} 
\bigg{|} \dfrac{i\alpha' t}{\pi R_{i_{c}}^{2}}\right)
~\nonumber\\
&\times & {\mathop{\lim }_{t\to \infty}}
\left(e^{3t}+\left[21-p +{\rm Tr}
\left(Q^\dagger_{(n=1)1} 
Q_{(n=1)2}\right)\right]e^{-t}\right) \bigg{\}}.
\label{tg}
\end{eqnarray} 
According to the negative mass squared of the tachyon, 
the divergent part in the last line exhibits 
exchange of the tachyonic state. 
The last bracket in Eq. (3.11) clarifies that 
in the twisted sector the $\mathbb{Z}_2$
projection extremely damps the long-range force.
This is due to the fact that this
projection modified the zero-point energy of the
Hamiltonian of this sector.

In fact, the twisted spectrum of closed string 
does not have any massless state, but contains the 
tachyonic state with a modified imaginary mass.
Therefore, the vanishing long-range force in this 
sector is an expected result. However, we
calculated this force to find the damping form of 
it and the divergence 
form for the tachyon exchange.

We should also calculate the long-time behavior 
of the interaction amplitude in the untwisted sector.
By considering the following limit in the 
26-dimensional spacetime
\begin{eqnarray}
&~& {\mathop{\lim }_{t\to \infty}}
e^{4t} \prod_{n=1}^\infty \bigg{(} 
\det[\mathbf{1}-Q^\dagger_{(n)1}
Q_{(n)2}e^{-4nt}]^{-1}~
\left(1- e^{-4nt}\right)^{p-23}\bigg{)}
\nonumber\\
&~& \longrightarrow
e^{4t}+ 23-p +{\rm Tr}\left(Q^\dagger_{(n=1)1} 
Q_{(n=1)2}\right),
\end{eqnarray}
the long-range force of the untwisted sector
takes the form  
\begin{eqnarray}
\mathcal{A}_{\rm long-range}^{\mathcal{U}} 
&=&\frac{T_p^2\alpha'V_{{p+1}}}{4(2\pi)^{25-p}}
\frac{\prod_{n=1}^{\infty}[\det(M^\dagger_{(n)1} 
M_{(n)2})]^{-1}}{\sqrt{\det{(U_1/2)}\det{(U_2/2)}}}
\int_{0}^{\infty}dt\bigg{\{}\Big{(}
\sqrt{\frac{\pi}{\alpha' t}}\Big{)}^{d_{I_n}}
~\nonumber\\
&\times& [\det \mathcal{Z}(t)]^{-1/2}
\exp\left( {-\frac{1}{4\alpha't}
\sum_{i}{\left(y_{1}^{I_n}-y_{2}^{I_n}\right)^2}}
\right)\prod_{I_{c}}\Theta_{3} 
\left(\dfrac{y_1^{I_{c}}-y_2^{I_{c}}}{2\pi R_{I_{c}}} 
\bigg{|} \dfrac{i\alpha' t}{\pi R_{I_{c}}^{2}}\right)
~\nonumber\\
&\times & \left( {\mathop{\lim }_{t\to \infty}}
e^{4t}+ 23-p +{\rm Tr}
\left(Q^\dagger_{(n=1)1} 
Q_{(n=1)2}\right)\right) \bigg{\}}.
\label{tg}
\end{eqnarray}
Again the divergent part represents the exchange of
the tachyon state, and the remainder indicates the 
long-range force.

The amplitudes (3.11) and (3.13) demonstrate that
the orbifold projection does not
deform the total long-range force. In addition, 
this projection imposed the divergence 
$e^{3t}$ as the contribution of the 
tachyon exchange in the twisted sector.
Besides, these amplitudes reveal that the compactification of the 
branes directions does not have any role in the long-range 
force.

According to Eqs. (2.14) the matrices 
$Q_{(n)1}$ and $Q_{(n)2}$ contain 
$2(p+1)(2p+1)$ parameters 
\begin{eqnarray}
\{\omega_{(l)\alpha \beta}, F_{(l)\alpha \beta},
B_{(l)\alpha \beta},
U_{(l)\alpha \beta} |\alpha , \beta =0,1,\ldots , p\},
\nonumber
\end{eqnarray}
with $l=1,2$ for the first and second interacting branes.
By adjusting these parameters we can receive  
\begin{eqnarray}
23-p +{\rm Tr}\left(Q^\dagger_{(n=1)1} 
Q_{(n=1)2}\right)=0,
\end{eqnarray}
and hence, we acquire a vanishing total long-range force.
In fact, for the two D0-branes there are only two parameters 
$U_{(1)00}$ and $U_{(2)00}$, thus this equation 
is not satisfied. However, for the systems with 
$p \geq 1$ there are enough parameters for 
satisfying this equation.
For example, consider two parallel D1-branes.
For simplification let $\omega_{(1)01}=\omega_{(2)01}=0$,
therefore Eq. (3.14) is decomposed to the following equations  
\begin{eqnarray}
&~&U_{(1)11} U_{(2)00}U_{(2)11}-U_{(1)00}U_{(1)11}
U_{(2)11}+U_{(1)00}U_{(1)11}U_{(2)00}
\nonumber\\
&~& -U_{(1)00}U_{(2)00}U_{(2)11}
+4U_{(1)11}-4U_{(2)11}+4U_{(2)00}-4U_{(1)00}
\nonumber\\
&~&-4U_{(1)11}\mathcal{F}^{2}_{(2)01} 
+4U_{(2)11}\mathcal{F}^{2}_{(1)01}
-4U_{(2)00}\mathcal{F}^{2}_{(1)01}+4U_{(1)00}\mathcal{F}^{2}_{(2)01}
\nonumber\\
&~&-U_{(1)11} U^{2}_{(2)01}+U_{(2)11} U^{2}_{(1)01}
-U_{(2)00} U^{2}_{(1)01}+U_{(1)00} U^{2}_{(2)01}= 0~,
\nonumber\\
\nonumber\\
&~&-12U_{(1)11}U_{(2)11}-48\mathcal{F}^{2}_{(1)01}
\mathcal{F}^{2}_{(2)01}-4U_{(1)01}U_{(2)01}
-16\mathcal{F}_{(1)01}\mathcal{F}_{(2)01}
\nonumber\\
&~&-12U^{2}_{(1)01}\mathcal{F}^{2}_{(2)01}
-12U^{2}_{(2)01}\mathcal{F}^{2}_{(1)01}
-12U_{(1)00}U_{(2)00}+10U_{(1)11}U_{(2)00}
\nonumber\\
&~& +10U_{(1)00}U_{(2)11}+3U^{2}_{(1)01}U_{(2)00}U_{(2)11}
+3U^{2}_{(2)01}U_{(1)00}U_{(1)11}+10U^{2}_{(2)01}
\nonumber\\
&~&+10U^{2}_{(1)01}+40\mathcal{F}^{2}_{(2)01}
+40\mathcal{F}^{2}_{(1)01}-10U_{(1)00}U_{(1)11}
-10U_{(2)00}U_{(2)11}
\nonumber\\
&~&-3U^{2}_{(1)01}U^{2}_{(2)01}+12\mathcal{F}^{2}_{(1)01}
U_{(2)00}U_{(2)11}+12\mathcal{F}^{2}_{(2)01}U_{(1)00}U_{(1)11}
\nonumber\\
&~& -3U_{(1)00}U_{(1)11}U_{(2)00}U_{(2)11}-48= 0~.
\end{eqnarray}
 
\section{Conclusions}

We constructed the boundary states, associated with a 
non-stationary fractional-wrapped D$p$-brane,
in the presence of the Kalb-Ramond 
background field, an internal $U(1)$ 
gauge potential and an internal open string tachyon 
field in the twisted and untwisted sectors of 
the orbifold projection. 

We observed that the 
emitted closed strings cannot wrap around the 
compact directions which are perpendicular to the brane.
In addition, wrapping of them around the compact 
directions of the brane is controlled by the tachyon matrix.
Besides, each emitted closed string possesses a
momentum along the worldvolume of the brane.
This momentum depends on the position of the 
closed string center-of-mass,
its winding numbers, and the parameters of the 
setup. This noticeable result 
clarifies that the background fields, accompanied 
by the toroidal compactification and linear 
and angular velocities of the brane, induce a 
marvelous potential on the emitted closed string.

For both twisted and untwisted sectors the interaction amplitudes 
of two dynamical fractional-wrapped D$p$-branes, in the 
above-mentioned setup, were obtained. The multiplicity of the
parameters designed a generalized 
amplitude. The strength of the interaction accurately 
is adjustable via these parameters to any desirable value.

From the total interaction amplitude the total long-range force 
was extracted. The  
long-range force only originates from the untwisted sector.
That is, the orbifold directions quenches 
the contribution of the massless states to this 
interaction. By a specific adjustment of the 
parameters we can eliminate the long-range force.

  
\end{document}